\documentclass[traditabstract]{aa}

\usepackage{graphicx}
\usepackage{txfonts}
\usepackage{natbib}
\usepackage{xspace}

\usepackage{url}

\bibpunct{(}{)}{;}{a}{}{,}

\begin{document}

\title{COSMOGRAIL: the COSmological MOnitoring of \\GRAvItational Lenses\thanks{Based on observations made with the 1.2-m Swiss Euler telescope (La Silla, Chile), the 1.3-m SMARTS telescope (Las Campanas, Chile), and the 1.2-m Mercator Telescope. Mercator is operated on the island of La Palma by the Flemish Community, at the Spanish Observatorio del Roque de los Muchachos of the Instituto de Astrof\'isica de Canarias.}\fnmsep\thanks{The light curves are available at the CDS via anonymous ftp to cdsarc.u-strasbg.fr (130.79.128.5) or via http://cdsarc.u-strasbg.fr/viz-bin/qcat?J/A+A, and on http://www.cosmograil.org.}}

\subtitle{XIII. Time delays and 9-yr optical monitoring of \\the lensed quasar RX~J1131$-$1231}

\titlerunning{COSMOGRAIL XIII -- Time delays of RX~J1131$-$1231}

\author{
M. Tewes\inst{\ref{epfl}} \and
F. Courbin\inst{\ref{epfl}} \and
G. Meylan\inst{\ref{epfl}} \and
C. S. Kochanek\inst{\ref{ohio}, \ref{ohiocosmo}} \and
E. Eulaers\inst{\ref{liege}} \and
N. Cantale\inst{\ref{epfl}} \and
A. M. Mosquera\inst{\ref{ohio}, \ref{ohiocosmo}} \and
P. Magain\inst{\ref{liege}} \and
H. Van Winckel\inst{\ref{leuven}} \and
D. Sluse\inst{\ref{bonn}} \and
G. Cataldi\inst{\ref{epfl}} \and
D. V\"or\"os\inst{\ref{epfl}} \and
S. Dye\inst{\ref{nottingham}}
}

\authorrunning{Tewes et al.}

\institute{
Laboratoire d'astrophysique, Ecole Polytechnique F\'ed\'erale de Lausanne (EPFL), Observatoire de Sauverny, 1290 Versoix, Switzerland, \email{malte.tewes@epfl.ch} \label{epfl}
\and
Department of Astronomy, The Ohio State University, 140 West 18th Av., Columbus, OH 43210, USA \label{ohio}
\and
Center for Cosmology and Astroparticle Physics, The Ohio State University, 191 West Woodruff Av., Columbus, OH 43210, USA \label{ohiocosmo}
\and
Institut d'Astrophysique et de G\' eophysique, Universit\' e de Li\`ege, All\'ee du 6 Ao\^ut, 17, 4000 Sart Tilman, Li\`ege 1, Belgium \label{liege}
\and
Instituut voor Sterrenkunde, Katholieke Universiteit Leuven, Celestijnenlaan 200B, 3001 Heverlee, Belgium  \label{leuven}
\and
Argelander-Institut f\"ur Astronomie, Auf dem H\"ugel 71, 53121, Bonn, Germany \label{bonn}
\and
School of Physics and Astronomy, University of Nottingham, University Park, Nottingham NG7 2RD, UK \label{nottingham}
}

\date{Received 8 September 2012 / Accepted 31 May 2013}

\abstract{
We present the results from nine years of optically monitoring the gravitationally lensed $z_{\mathrm{QSO}}=0.658$ quasar RX~J1131$-$1231. The R-band light curves of the four individual images of the quasar were obtained using deconvolution photometry for a total of 707 epochs. Several sharp quasar variability features strongly constrain the time delays between the quasar images. Using three different numerical techniques, we measure these delays for all possible pairs of quasar images while always processing the four light curves simultaneously.
For all three methods, the delays between the three close images A, B, and C are compatible with being 0, while we measure the delay of image D to be 91 days, with a fractional uncertainty of $1.5\%$ ($1\sigma$), including systematic errors. Our analysis of random and systematic errors accounts in a realistic way for the observed quasar variability, fluctuating microlensing magnification over a broad range of temporal scales, noise properties, and seasonal gaps. Finally, we find that our time-delay measurement methods yield compatible results when applied to subsets of the data.
}

\keywords{Gravitational lensing: strong -- quasars: individual: RX~J1131$-$1231 -- cosmological parameters}

\maketitle
\section{Introduction}

Using the time delays between multiple images of gravitationally lensed sources to measure cosmological distances \citep{Refsdal:1964vh} has several advantages: there is no need for any primary or secondary calibrator, and there are no effects from the intergalactic or interstellar medium. The method, originally proposed for gravitationally lensed supernovae, has so far exclusively been applied to quasars lensed in most cases by individual massive galaxies. Exceptions are SDSS J1004$+$4112 and J1029$+$2623, two quasars lensed by galaxy clusters, with long time delays \citep{Fohlmeister:2008df, Fohlmeister:2013ji}. The quasar lens time-delay method is now recognized as a tool that complements other cosmological probes, in particular for constraining $H_{\rm{0}}$ as well as the dark-energy equation-of-state parameter, $w$ \citep[e.g.,][]{Suyu:2012tw, Linder:2011cs, Moustakas:2009wj}. In spite of its advantages, the method has long faced two severe limitations to its effectiveness in constraining cosmology. 

First, time delays between the gravitationally lensed images of a quasar are hard to measure. Some claimed time-delay measurements turned out to be erroneous \citep[see, for instance, the controversy around Q0957+561:][and references therein]{Vanderriest:1989uj, Press:1992jj, Schild:1997hf, Kundic:1997br}. Understandably, early light curves tended to be short and sparse, often too short to clearly demonstrate that \emph{microlensing} variability was not interfering with their analysis. Microlensing is seen as an uncorrelated extrinsic variability in the quasar images, which results from the time-variable magnification created by stars in the lensing galaxy \citep[e.g.,][]{Chang:1979dk, Schmidt:2010ce}. In the best cases, light curves spanned a few years \citep[see, e.g.,][]{Wyrzykowski:2003wv, Hjorth:2002gz, Burud:2002gb, Burud:2002ge}. One consequence is that the numerical methods used to measure time delays from these light curves were exceedingly ``optimistic'' in their assumptions about extrinsic variability. More recent measurements with better data frequently yielded delays inconsistent with the error estimates of the earlier measurements.

Second, given the measured time delays and lens and quasar image astrometry, there is a famed degeneracy between the \emph{time-delay distance}, which is a scale parameter inversely proportional to $H_{\rm{0}}$, and the spatial distribution of mass responsible for the strong-lensing phenomenon. The delays constrain only a combination of this time-delay distance and the surface density of the lens near the images \citep{Kochanek:2002in}. This can be overcome with independent constraints on the structure of the lens. \citet{Suyu:2009ig, Suyu:2010fq} convincingly showed that it is possible to control the effects of model degeneracies for B1608+656 \citep[][CLASS survey]{Myers:1995gz}, a quadruply imaged quasar with accurate radio time delays \citep{Fassnacht:2002ig}. To do this, the authors combined (1) detailed HST images of the lensed quasar host galaxy, (2) a velocity-dispersion measurement of the lens galaxy, and (3) information about the contribution of intervening galaxies along the line of sight, from galaxy number counts calibrated with numerical simulations.

In parallel to the advances in lens modeling, the observational situation has impressively evolved as well. Two observational groups, the COSmological MOnitoring of GRAvItational Lenses (COSMOGRAIL) and \citet{Kochanek:2006fp}, have been intensely monitoring $\approx 20$ lenses for roughly ten years. In 2010, our two groups decided to merge their observational efforts, with the COSMOGRAIL group focusing on the analysis of time delays and the \citet{Kochanek:2006fp} group focusing on the analysis of microlensing. While preliminary results have been published both before and after this merger \citep{Kochanek:2006fp,Vuissoz:2007du,Vuissoz:2008bu,Morgan:2008fc,Morgan:2008jb}, exquisite data spanning almost a decade of continuous observation are now being released, for instance for the quadruply imaged quasar HE~0435-1223 \citep{Courbin:2011bl, Blackburne:2011ue}.

In this paper, we present nine years of optical monitoring of the quadruply imaged quasar RX~J1131$-$1231 \citep{Sluse:2003cw}, and measure its time delays with the techniques of \citet{pycs}.
RX~J1131$-$1231 is one of the most spectacular lenses of our sample. The redshift of the lensing galaxy is $z_{\mathrm{lens}} = 0.295$, while the quasar is at $z_{\mathrm{QSO}} = 0.658$. This low quasar redshift means (1) that the photometric variations are fast, numerous, and \emph{strong} because it is a lower-luminosity quasar \citep[see, e.g.,][]{MacLeod:2010ex}, and (2) that the host galaxy of the lensed quasar is seen as a full Einstein ring with many spatially resolved structures in HST images. Similarly, the lensing galaxy is sufficiently bright to allow a precise measurement of its velocity dispersion and possibly of its velocity dispersion profile. These characteristics facilitate both the time-delay measurement and the lens modeling. The latter, with state-of-the-art inferences of cosmological constraints based on our time-delay measurements of RX~J1131$-$1231, are presented in \citet{suyu1131}.

Our observations of RX~J1131$-$1231 and their reduction are described in Sections~\ref{obs} and \ref{reduc}, while the light curves are presented in Section \ref{lc}. In Section~\ref{timedelay}, we apply three different curve-shifting techniques to the light curves and infer our best measurements of the delays along with realistic random and systematic error bars. Our results are summarized in Section~\ref{conclusions}. 

\section{Observations}
\label{obs}

We have been monitoring the quadruply lensed quasar RX~J1131$-$1231 (J2000: $11^{\mathrm{h}}31^{\mathrm{m}}52^{\mathrm{s}}$, $-12\degr31\arcmin59\arcsec$) since December 2003 with three different telescopes in the R band ($\sim 600$ - 720 nm). Table \ref{monitoring} summarizes the observational strategy and instrumental characteristics. The light curves presented in this paper cover nine observing seasons (2004 - 2012, see Fig. \ref{fig_lc}), with 707 monitoring epochs in total. The average sampling within the seasons is 2.9 days, and the median sampling is 2.0 days. The mean seasonal gap in the combined light curves is $132$ days with a standard deviation of two weeks.

\begin{table*}[htdp]
\caption{Overview of our optical monitoring of RX~J1131$-$1231}
\begin{center}
\begin{tabular}{lllcclrr}
\hline
\hline
Telescope & Location & Instrument & Pixel scale & Field & Exp. time & Epochs & Sampling \\
\hline
Euler 1.2 m & ESO La Silla, Chile   & C2 &  0\hspace{1.5 pt}\farcs344 & $11\arcmin \times 11\arcmin$ & $5 \times 360$ s & 265 & 5.0 (4.0)\\
Euler 1.2 m & ESO La Silla, Chile   & EulerCAM & 0\hspace{1.5 pt}\farcs215 & $14\arcmin \times 14\arcmin$ & $5 \times 360$ s & 76 & 5.4 (4.1)\\
SMARTS 1.3 m & CTIO, Chile & ANDICAM & 0\hspace{1.5 pt}\farcs369 & $6.1\arcmin \times 6.1\arcmin$ & $3 \times 300$ s & 288 & 6.1 (5.1)\\
Mercator 1.2 m & La Palma, Canary Islands, Spain  & MEROPE & 0\hspace{1.5 pt}\farcs193 & $6.5\arcmin \times 6.5\arcmin$ & $5 \times 360$ s & 78 & 8.9 (4.5)\\
\hline
\end{tabular}
\tablefoot{The column ``Exp. time'' indicates the number of dithered exposures \emph{per epoch} and their individual exposure times. One epoch corresponds to one data point in each light curve. The temporal sampling of the observations is given as the average (median) number of days between consecutive epochs, excluding the seasonal gaps.}
\end{center}
\label{monitoring}
\end{table*}

\begin{figure}[tbp]
\resizebox{\hsize}{!}{\includegraphics{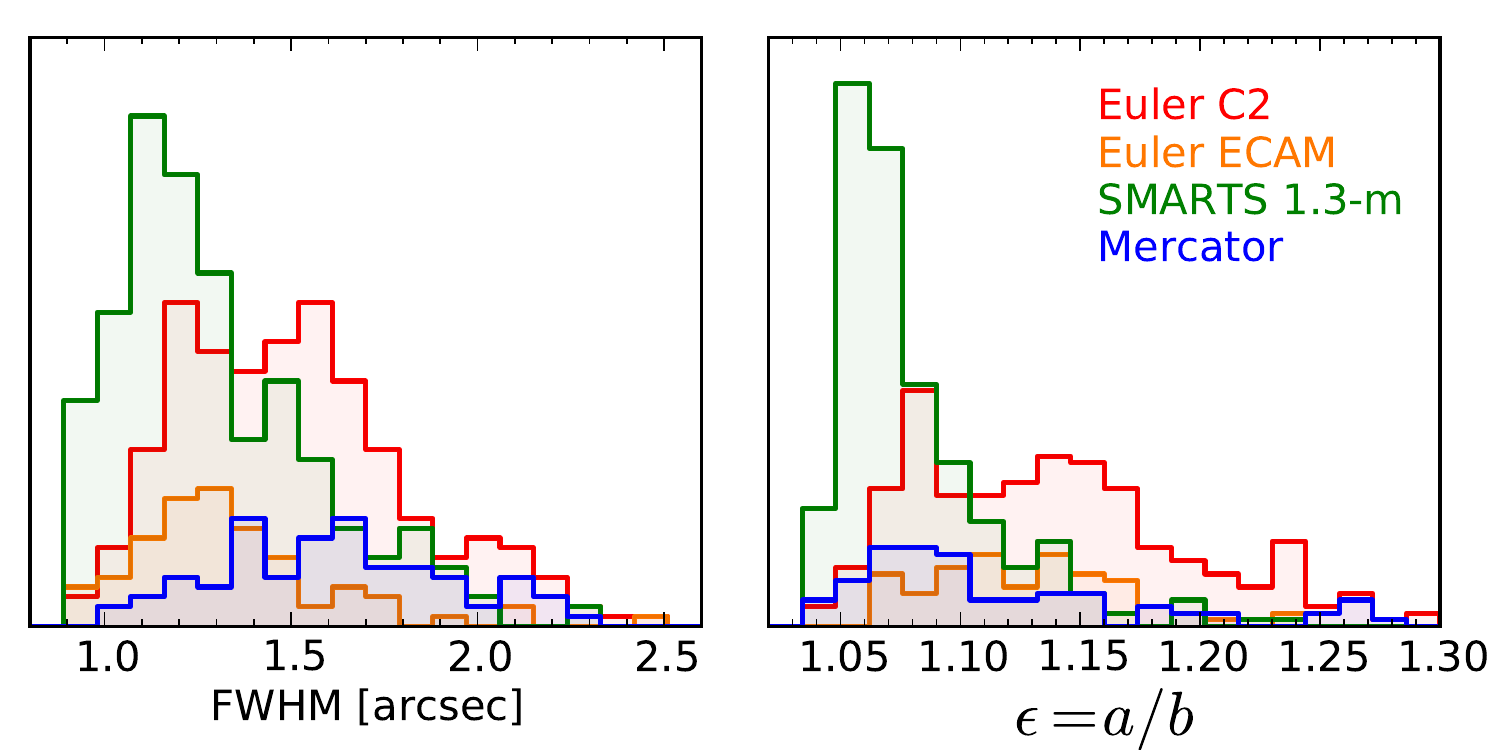}}
\caption{Distributions of stellar FWHM and elongation $\epsilon$ of all images contributing to the light curves, by telescope.}
\label{fig_seeing}
\end{figure}

The majority of the measurements for this southern target came from the Swiss 1.2-m Euler telescope and the Small \& Moderate Aperture Research Telescope System (SMARTS) 1.3-m telescope, both located in Chile under equally good atmospheric conditions. Fig. \ref{fig_seeing} shows the distributions of the stellar full width at half maximum (FWHM) and the elongation $\epsilon$ of each observing epoch, as measured by SExtractor \citep{Bertin:2010wt} using field stars. The SMARTS 1.3-m telescope is guided, in contrast to Euler and Mercator, which are solely tracking. This accounts in part for the broad elongation distribution of the Euler and Mercator images.

The original imaging instrument C2 of the Euler telescope was replaced in September 2010 by EulerCAM, a liquid-nitrogen cooled 4k $\times$ 4k e2v 231-84 CCD yielding a pixel scale of 0\hspace{1.5 pt}\farcs215. At the same time the focusing procedure was improved. Figure \ref{fig_seeing} shows that after two years of observations, images from EulerCAM tend to be statistically sharper than C2 images, and the FWHM is more similar to the SMARTS 1.3-m data. All images from the SMARTS 1.3-m telescope were obtained through the optical channel of the ANDICAM\footnote{\url{http://www.astronomy.ohio-state.edu/ANDICAM/}} camera \citep{andicam}. See \citet{Kochanek:2006fp} for details about the SMARTS data.

\begin{figure*}[htbp]
\resizebox{\hsize}{!}{\includegraphics{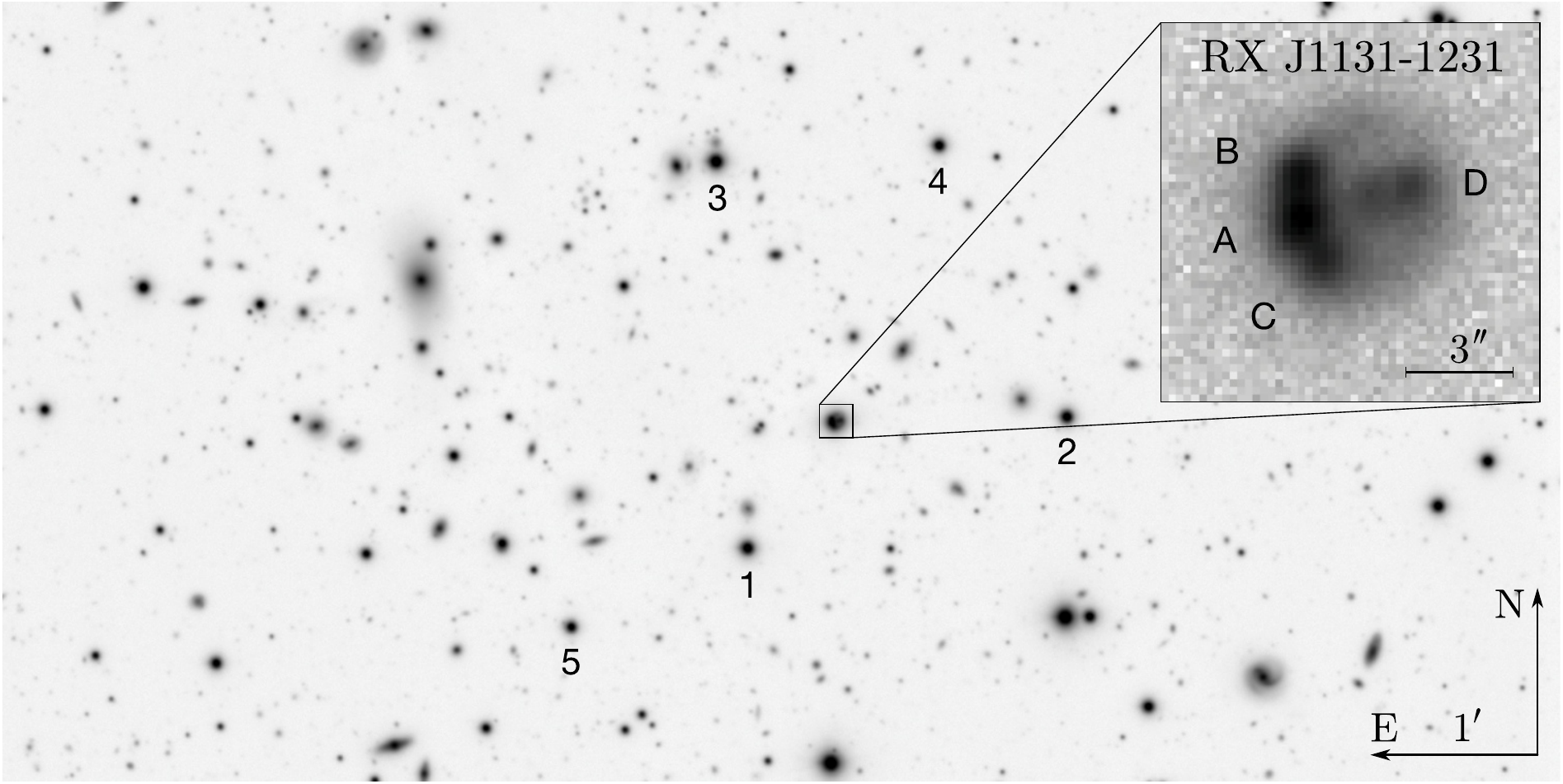}}
\caption{Part of the field of view of the Swiss 1.2-m Euler telescope around RX~J1131$-$1231. The wide-field image is a combination of 600 exposures of 360s each, corresponding to a total exposure time of 2.5 days. The inset shows a single 360s exposure from the \emph{EulerCAM} instrument under good conditions (FWHM$\sim$1\arcsec, $\epsilon = 0.1$, no Moon). All images are taken in the R band. We mainly use the stars labeled 1 to 4 to build a PSF model for each exposure. Stars 1 to 5 are used for the photometric calibration.
}
\label{fig_field}
\end{figure*}

Figure \ref{fig_field} shows the central part of a deep combination of Euler monitoring images, totaling 2.5 days of exposure. In the best individual exposures the four quasar images are well resolved; images A and B have the smallest separation of 1\farcs2.

\section{Image reduction}
\label{reduc}

In this section we describe the reduction procedure leading to the light curves of the multiple quasar images of RX~J1131$-$1231. This same procedure will homogeneously be applied to other lenses monitored by COSMOGRAIL and SMARTS, hence we provide a general description of our methods.

All exposures are corrected for bias/readout effects and flat fielded using standard methods. This prereduction is performed individually for each telescope and instrument. For Euler and Mercator we have developed custom \verb+python+ pipelines that enable efficient human inspection and validation of each step. Particular attention is given to a semi-automated selection and combination of sky-flats, which is controlled through a simple graphical interface highlighting the temporal evolution of instrumental contaminations such as dust. Using difference images, we check that all sources in the flatfield exposures are adequately masked and that the contamination does not evolve significantly within sets of flatfields that are stacked as master-flats. The period over which we stack flatfields can be from single days to several weeks. Some exposures of the Mercator telescope are prereduced using superflats, combinations of masked science exposures obtained in a single night. We have developed variants of these pipelines adapted to the particularities of other COSMOGRAIL telescopes.

From here on, all images are processed by a single deconvolution photometry software package. The principle ideas for this reduction procedure are the same as employed for previously published COSMOGRAIL light curves \citep[][]{Vuissoz:2007du,Vuissoz:2008bu,Courbin:2011bl}. Over the years we have reworked the steps, and implemented the procedure in the form of a \verb+python+ pipeline linked to a relational database, containing one entry per exposure.

\subsection{PSF construction}
\label{psf}

We first build a conservatively smooth sky model, obtained using SExtractor, for each exposure. This sky model is not critical, as the subsequent photometry procedure will fit for any residual sky level around all sources of interest.

We then align the images, separately for each instrument, and individually estimate the point spread function (PSF) of each exposure. This is done by fitting several field stars using a common model, composed of (i) a simply parametrized profile and (ii) a regularized fine pixel array. The details of this procedure, which is part of a general-purpose deconvolution package based on ideas reported in \citet{Magain:1998ko}, are described in Cantale et al. (in prep). For most of the exposures of RX~J1131$-$1231, we used the stars labeled 1 - 4 in Fig. \ref{fig_field} to build the PSF model. The pipeline allows us to easily explore and compare different choices of PSF-stars. The final choice of stars is empirically selected to yield the least scatter in the final quasar light curves. For RX~J1131$-$1231 the situation is close to optimal, since (1) the stars 1 - 4 are bright but still in the linear regime of the CCDs, (2) they surround the lens at modest angular distances, and (3) all stellar companions or background objects can be identified and masked, yielding clean fitting residuals. Cosmic-ray hits and CCD artifacts are automatically masked using a variant of the L.A. Cosmic algorithm \citep{vanDokkum:2001go}. We visually supervise the PSF construction and manually adapt the selection of PSF-stars for problematic frames. The construction of the PSF models is the predominant technical issue affecting the final quality of the light curves.

\subsection{Photometric normalization of the exposures}
\label{norm}

We next compute a multiplicative normalization coefficient for each exposure, based on the photometry of field stars. These coefficients will constitute the reference for the differential photometry of the quasar images. 

In a first step, this computation is made separately for each instrument. We measure the instrumental fluxes $N_{\star ij}$ (in photons) of star $i$ in each exposure $j$, by fitting the $\mathrm{PSF}_{j}$ of this exposure to each star, leaving only the fluxes, astrometric shifts, and background levels of each star as free parameters. For each star and exposure, we then compute an individual calibration coefficient
\begin{equation}
\label{coeffindiv}
c'_{ij} = \frac{\mathrm{med}_{j}(N_{\star ij})}{N_{\star ij}},
\end{equation}
where $\mathrm{med}_{j}$ denotes the median over all exposures of this instrument. Then, for each exposure $j$, we obtain the normalization coefficient through
\begin{equation}
\label{coeff}
c_{j} = \mathrm{med}_{i}(c'_{ij}).
\end{equation}
The advantages of this simple and fast method is that it does not select a single exposure as the ``reference'' for all stars. For each instrument, the distribution of these coefficients $c_{j}$ is highly unimodal, because the exposure time is kept constant.

To select the normalization stars we first inspect the normalized stellar light curves for any anomalies or variability on scales of months or years, and, if required, we iteratively repeat the coefficient computation with a refined selection of stars. In terms of the smoothness of the final quasar light curves, the best results are generally obtained by computing these coefficients from a few well-exposed stars with a similar and high signal-to-noise ratio, as opposed to using larger numbers of noisier stars. Furthermore, the ideal normalization stars should be of similar color to the quasar. This minimizes potential differential effects due to varying atmospheric conditions given the relatively broad R-band filters used for the monitoring.

In a second step, we rescale the coefficients of each instrument so that a star whose color is closest to that of the quasar receives the same normalized median flux across all instruments. Residual adjustments to this normalization between instruments are performed later, using the quasar light curves in temporal regions where the data from different telescopes overlap.

\subsection{Photometry of the quasar images}

We obtain deblended light curves of the quasar images by \emph{MCS deconvolution photometry}, following \citet{Magain:1998ko}. This algorithm fits a single model consisting of (1) some number of point sources and (2) a fine pixel array (hereafter the \emph{pixel channel}) simultaneously to all exposures. For RX~J1131$-$1231, four point sources model the quasar images, while the pixel channel represents both the lensing galaxy and the lensed host galaxy (Einstein ring). This model is scaled by the normalization coefficients from Equation \ref{coeff} and convolved by the exposure-specific PSF, before it is fit to the data. By construction, the relative astrometry of the four point sources and the pixel channel are common to all exposures. Only the fluxes of the point sources, the absolute astrometric shift, and a flat residual sky level around the lens are let free to vary between exposures. Iteratively, all these parameters are optimized together with the structure of the regularized pixel channel to minimize a single global $\chi^2$. This algorithm was previously successfully applied in the discovery paper of RX~J1131$-$1231 \citep{Sluse:2003cw}, as well as in past COSMOGRAIL publications \citep[e.g.,][]{Eigenbrod:2007fk, Vuissoz:2008bu, Courbin:2011bl, Sluse:2012ir} or for similar monitoring data \citep{Burud:2002gb, Hjorth:2002gz, Jakobsson:2005kw,Morgan:2012cl}. The light curves of the quasar images are a direct output of this procedure.

Errors in the astrometry or the structure of the pixel channel might degrade or bias the photometry of the quasar images. We observe that the influence of the astrometry on the light curves is weak. Even for  bright quasar images, alterations of the relative position of the point sources by as much as 0\hspace{1.5 pt}\farcs05 do not significantly modify the light curves. Larger errors tend first to smoothly bias the magnitude measurements, before introducing additional scatter. To obtain light curves for gravitational lenses with image separations on the order of the resolution of the best monitoring data, we find no difference between using the tight astrometric constraints from HST images \citep{Morgan:2006uw, Chantry:2010cl} or letting the deconvolution algorithm freely optimize the astrometry of the model.

We reach similar conclusions for the impact of the pixel channel and its mandatory regularization. Simple numerical experiments show that to first order, the effect of different plausible solutions is similar to very small additive shifts to the flux of the quasar light curves, with only marginally perceptible effects even for faint quasar images. In practice, we systematically constrain this pixel channel as well as the point source astrometry using only a subset of images with the best resolution.

\subsection{Photometric error estimation}

In this section we describe how we compute a rather formal best case error estimate for the photometry of each quasar image in each exposure. 
The normalized flux of a quasar image in a given exposure can be written $f_{\star} = N_{\star} \cdot c$, where $N_{\star}$ is the measured number of photons, and $c$ is the normalization coefficient from Equation \ref{coeff}. We assume that the random error on $f_{\star}$ has two independent sources: (i) the shot noise $\sigma_{N_{\star}}$ of the quasar image itself, and (ii) the noise $\sigma_c$ of the normalization coefficient as computed in Section \ref{norm}.

We compute $\sigma_{N_{\star}}$ following the standard ``CCD equation'' \citep[see e.g.][chap. 4.4]{Howell:2006vm}
\begin{equation}
\sigma_{N_{\star}} = \sqrt{N_{\star} + n_{\mathrm{pix}}\cdot(S + R^2)},
\label{ccd}
\end{equation}
where $S$ is the sky level, $R$ is the CCD read noise (both in photons per pixel), and $n_{\mathrm{pix}}$ is the number of (equivalent) pixels of the software aperture. The dark current is negligible in our images. The MCS deconvolution of point sources corresponds to PSF fitting, and following \citet[][section 6.5.1]{Heyer:2004uv}, we use for $n_{\mathrm{pix}}$ the reciprocal of the PSF \emph{sharpness},
\begin{equation}
\frac{1}{n_{\mathrm{pix}}} = \sum_{l,m} (\mathrm{PSF}_{lm})^2,
\label{sharpness}
\end{equation}
where $\mathrm{PSF}_{lm}$ is the fraction of light in the total PSF at pixel $lm$. In this simple best case computation, we assume that the PSF is perfectly known, and that the sky level is well determined by the fitting procedure. 
We do not take into account the surface brightness of the lens and host galaxies, which is generally negligible with respect to the sky level. Furthermore, we do not compute noise terms that are due to the blending with other point sources.

As a sanity check, we tested our deconvolution photometry algorithm on simulated point source images created with \verb+SkyMaker+ \citep{Bertin:2009wz} for various realistic signal-to-noise ratios, elongations, and seeing conditions. Using two or more well-exposed stars to build the PSF, the statistical scatter obtained by our photometry pipeline is only marginally larger than the best-case shot noise computed through the above procedure. Using similar synthetic images, we also verified that deconvolving two partially blended sources of different fluxes does not significantly bias their measured flux difference in any direction.

To estimate the random uncertainty of the normalization coefficient, we use the standard error of the mean (SEM) of the individual coefficients obtained from the $n$ normalization stars in the exposure (see Equation \ref{coeffindiv})
\begin{equation}
\sigma_{c} = \frac{s}{\sqrt{n}} \quad \textrm{where} \quad s^2 = \frac{1}{n-1} \sum_{i=1}^{n}(c'_i - \bar{c'})^2.
\end{equation}
Finally, we obtain the error estimated $\sigma_{f_{\star} \, \mathrm{CCD}}$ on the normalized flux through
\begin{equation}
\label{sigmaformal}
\frac{\sigma_{f_{\star} \, \mathrm{CCD}}^2}{f_{\star}^2} = \frac{\sigma_{N_{\star}}^2}{N_{\star}^2} + \frac{\sigma_c^2}{c^2}.
\end{equation}
Note that for our COSMOGRAIL and SMARTS monitoring data, the shot noise term from the CCD equation dominates this error budget. The contribution from $\sigma_c$ is typically $\ll3\%$ of $\sigma_{f_{\star} \, \mathrm{CCD}}$. Such small calibration errors are important only when measuring very short delays, where they would introduce positively correlated noise into the curves.

\subsection{Combination of points per night}

In each monitoring night, we observe the lenses $m$ times ($m = 3$ or $5$) over $\approx 30$ minutes, yielding flux measurements $f_{\star k}$, with $k=1,\,..\, ,m$ for each quasar image. To reduce the CPU cost and to reject outliers, we bin these measurements by epochs, separately for each instrument and telescope, before measuring the time delays. We attribute to each epoch the \emph{medians} of the photometric measurements $f_{\star k}$ within that night, and the mean of the Heliocentric Julian Dates (HJD).

This approach also allows us to obtain an empirical photometric error estimate for each quasar image, using a measure of the spread of the image's $f_{\star k}$. For increased robustness against outliers, we quantify this spread using the median absolute deviation from the median, or \emph{MAD} \citep[][]{Hoaglin:1983tl},
\begin{equation}
\mathrm{MAD}(f_{\star k}) = \mathrm{med}(\, \left|\, f_{\star k} - \mathrm{med}(f_{\star k}) \, \right| \,).
\end{equation}
To estimate the usual $\sigma$ of an (assumed) Gaussian distribution, the MAD is rescaled as
\begin{equation}
\sigma_{f_{\star}\, \mathrm{MAD}} = 1.4826 \cdot  \mathrm{MAD}(f_{\star k}).
\end{equation}
These procedures give us two different error estimates for each epoch and quasar image: (1) the median of the errors estimated individually for each exposure $\mathrm{med}(\sigma_{f_{\star}\, \mathrm{CCD}})$, and (2) the more empirical $\sigma_{f_{\star}\, \mathrm{MAD}}$. We observe, as expected, that both error estimations are highly correlated, but also that the empirical $\sigma_{f_{\star}\, \mathrm{MAD}}$ is typically twice as high.

\begin{figure}[tb]
\resizebox{\hsize}{!}{\includegraphics{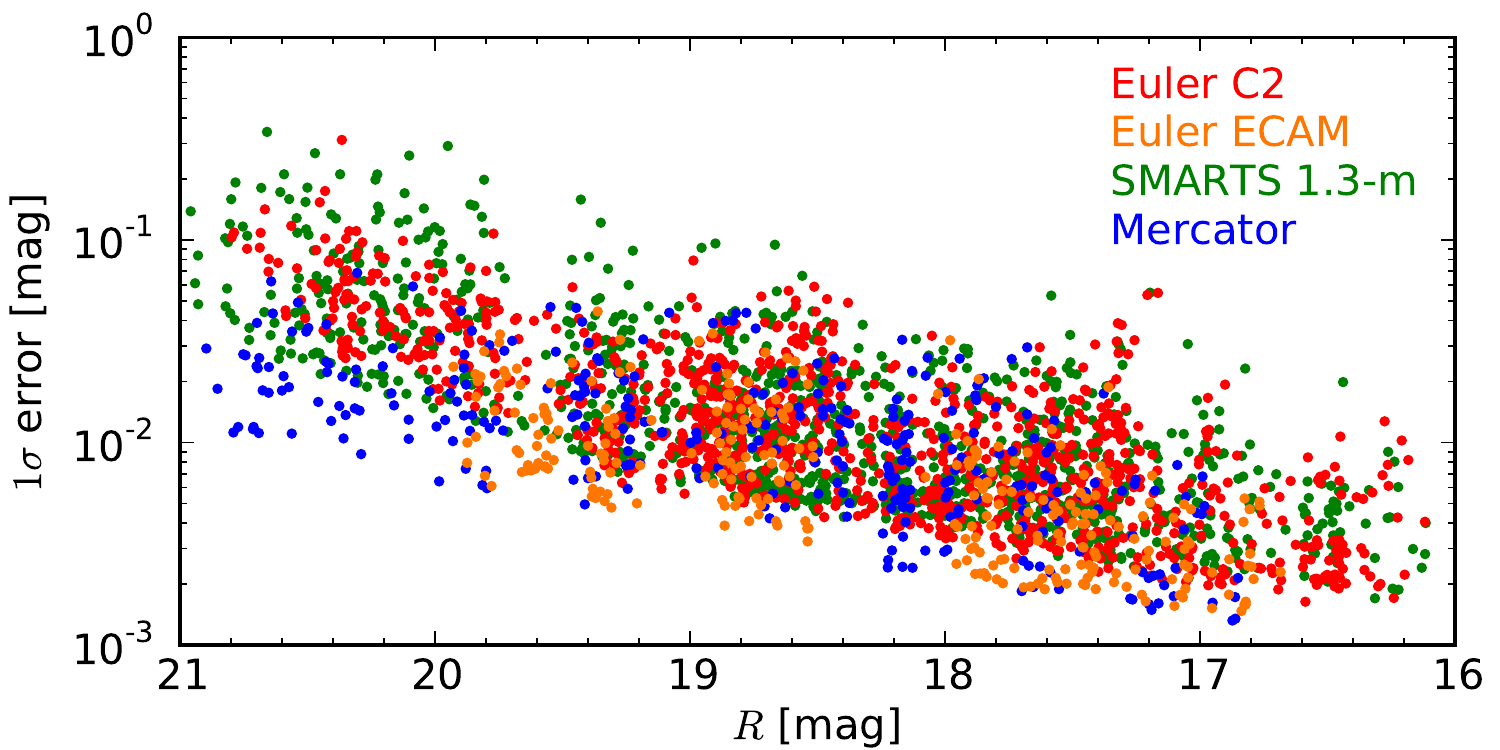}}
\caption{Photometric $1\sigma$ error estimates of each observing epoch as a function of approximate R-band magnitude. These errors are estimated following Equation \ref{sigma}.}
\label{fig_magerrs}
\end{figure}

The uncertainty we finally assign to each epoch and quasar image uses the higher of $\mathrm{med}(\sigma_{f_{\star \, \mathrm{CCD}}})$ and $\sigma_{f_{\star}\, \mathrm{MAD}}$. Indeed, a realistic error estimate cannot be lower than either of them. We divide this estimate by the square root of the number of exposures $m$ of the epoch, so that the final error estimate for the median photometric measurement on a given instrument and night is
\begin{equation}
\label{sigma}
\sigma_{\mathrm{med}(f_{\star})} = \frac{\mathrm{max}\left[ \mathrm{med}(\sigma_{f_{\star} \, \mathrm{CCD}}),\, \sigma_{f_{\star}\, \mathrm{MAD}}\right]}{\sqrt{m}}.
\end{equation}
Figure \ref{fig_magerrs} displays the distribution of the resulting error estimates as a function of R-band magnitude and instrument. The data points from all four quasar images of RX~J1131$-$1231 are shown, yielding a broad spread in magnitudes.

\begin{figure*}[t]
\resizebox{\hsize}{!}{\includegraphics{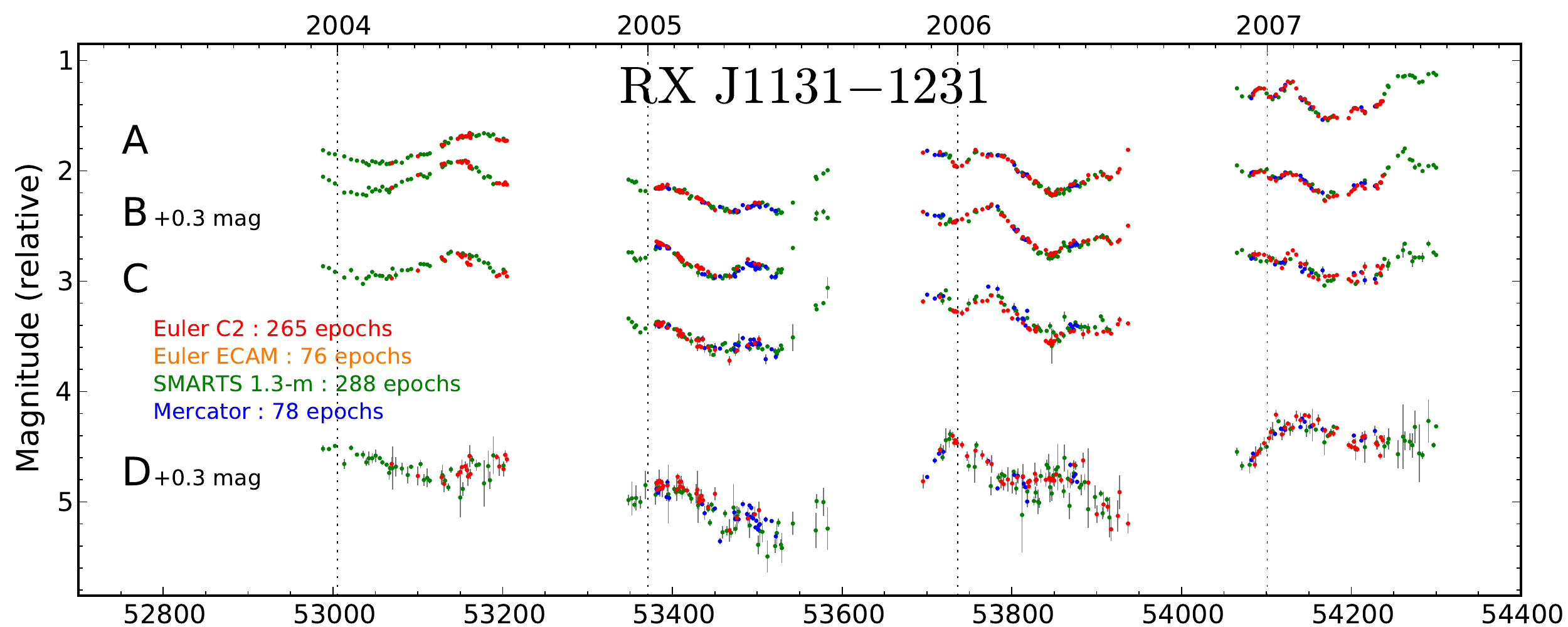}}
\resizebox{\hsize}{!}{\includegraphics{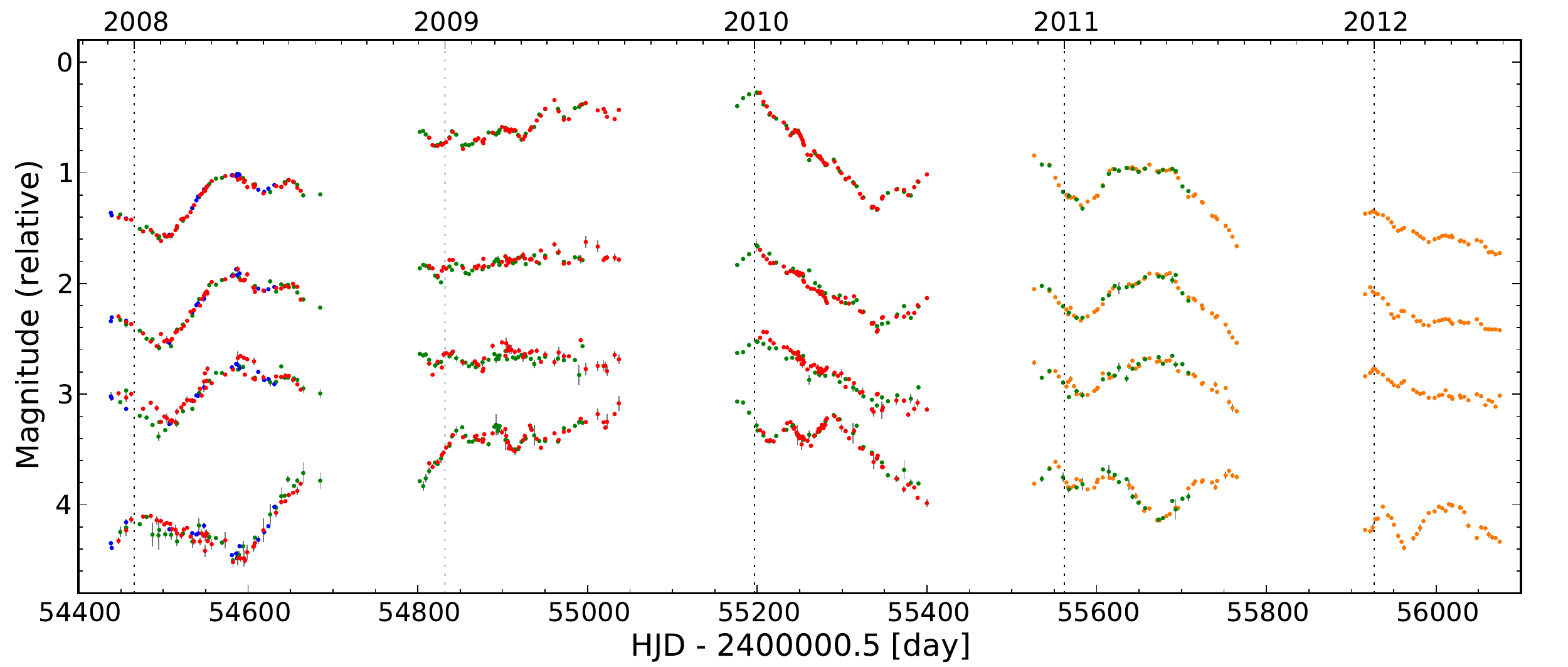}}
\caption{Optical monitoring of RX~J1131$-$1231, as obtained from deconvolution photometry. From top to bottom are shown the R-band light curves for the quasar images A, B, C, and D along with the $1\sigma$ photometric error bars. Colors encode the contributing instruments. Curves B and D have been shifted by +0.3 magnitudes for display purposes. The light curves are available in tabular form from the CDS and the COSMOGRAIL website.}
\label{fig_lc}
\end{figure*}

\subsection{Combination of telescopes}
\label{combitel}

The lens galaxy and the lensed images of the quasar host galaxy differ in color from the quasar itself. As a consequence, the different R filters and CCD response functions of the monitoring instruments might ``see'' these contaminating sources with slightly different amplitudes even with a perfect calibration of the quasar fluxes. This can result in small mismatches between the light curves obtained using different combinations of telescopes and instruments.

We correct our light curves for such small effects (occasionally as high as $10\%$ of the flux of the faintest quasar images) in this final step. We typically select the instrument with the longest or the highest-quality curve as a reference. For RX~J1131$-$1231, this is the SMARTS light curve. For each of the other instruments, we optimize additive magnitude and flux shifts (i.e., multiplicative and additive flux corrections) to minimize a \emph{dispersion measure} between each instrument's light curve and the reference light curve. We compute this dispersion following the curve-shifting technique presented in \citet{pycs}, but evaluate it between the light curves of different \emph{instruments}, instead of different quasar images. Provided that the colors of the quasar images are not differentially reddened by absorption in the lens galaxy, we optimize a single \emph{common} magnitude shift per instrument, and individual flux shifts for each quasar image and instrument. This is adequate for RX~J1131$-$1231, as previously suggested by \citet{Sluse:2007hw} and \citet{Chartas:2009bc, Chartas:2012dl}.

\section{Light curves of RX~J1131$-$1231}
\label{lc}

All these steps were applied to the COSMOGRAIL and SMARTS data for RX~J1131$-$1231. Fig. \ref{fig_lc} shows the final nine-year-long light curves of the four quasar images. These data are available in machine-readable form at the CDS and on the COSMOGRAIL website\footnote{\url{http://www.cosmograil.org}}. The light curves are dominated by intrinsic quasar variability, with some features on scales as short as a few weeks. It can be readily seen in the 2008 season for instance, that the delays between A, B, and C must be very short, while D is delayed by slightly less than 100 days. Intriguingly, looking only at the first season of A, B, and C, one might think that A is significantly delayed with respect to B and C. We attribute this discrepancy to microlensing variability, which manifestly changes the magnitude difference between the A and B images from the first to the second season. We discuss this ``event'' in more detail in Section \ref{subsections}. Prominent microlensing variability on long time-scales is ubiquitous because the flux ratios between the quasar images evolve by as much as a magnitude. These microlensing effects in RX~J1131$-$1231 have been analyzed in \citet{Morgan:2010bw} and \citet{Dai:2010im}.

Lastly, we observe that the photometric error estimates, obtained from equation \ref{sigma}, match the observed scatter well in the smooth sections of curves from the individual telescopes. They are certainly not conservatively large, but we stress that the scale of these error estimates has \emph{no} direct influence on the uncertainties that we compute for the time-delay measurements in the next Section. Our results are robust against a deliberate increase of these error estimates by up to a factor of 5.

\section{Time-delay estimation}
\label{timedelay}

In this section we infer the time delays of RX~J1131$-$1231 from the light curves shown in Fig. \ref{fig_lc}, closely following the curve-shifting and uncertainty evaluation procedures described in \citet{pycs}. We summarize the principal ideas below. A major difficulty and potential source of bias for curve-shifting methods is the presence of extrinsic variability in the light curves, on top of the intrinsic quasar variability common to all four images. The main source of the extrinsic signal is variable microlensing magnification due to the motions of the stars in the lens galaxy. As shown by \citet{Mosquera:2011uf}, microlensing can affect light curves over a broad range of time scales. For RX~J1131$-$1231, \citet{Mosquera:2011uf} estimated a time scale of $\approx 11$ years for the crossing of a stellar Einstein radius, and $\approx 3$ months for the source radius to cross a caustic. Other potential sources of extrinsic variability, such as variable quasar structure and spurious additive flux contaminating the photometry, are summarized in \citet{pycs}. In the present paper we do not aim to separate or analyze the extrinsic variability in physical terms. We simply consider it as a hindrance for the time-delay measurement.

The curve-shifting methods of \citet{pycs} try to minimize the bias due to such extrinsic variability in different ways. They all rely on iterative optimizations of time shifts of the four light curves, and yield self-consistent point estimates of the time delays between all six image pairs :
\begin{enumerate}
\item {\bf The free-knot spline technique} simultaneously fits one common \emph{intrinsic} spline for the quasar variability, and independent, smoother \emph{extrinsic} splines for the microlensing to the light curves. The curves are shifted in time to optimize this fit. This method is similar to the polynomial method of \citet{Kochanek:2006fp}.
\item {\bf The regression difference technique} shifts continuous regressions of the curves in time to minimize the \emph{variability} of the differences between them. This novel method does not involve an explicit model for the microlensing variability.
\item {\bf The dispersion-like technique}, inspired by \citet{Pelt:1996vy}, shifts the curves to minimize a measure of the \emph{dispersion} between the overlapping data points. This method has no explicit model for the common intrinsic variability of the quasar, but it includes polynomial models for the extrinsic variability.
\end{enumerate}

Using several independent algorithms allows us to cross-check our estimates for technique-dependent biases. Indeed, as important as the time-delay point estimates themselves is the reliable estimation of their uncertainties. For this we follow a Monte Carlo approach, by applying the curve-shifting techniques to a large number of synthetic light curves with \emph{known} time delays. These curves are drawn from a light curve model that mimics both the observed intrinsic and extrinsic variability of the real observations, randomizing only the unrecoverable short time-scale extrinsic variability. This latter correlated noise locally adapts its amplitude to the scatter of the observed data points. As a result, the synthetic curves are virtually indistinguishable from the real ones \citep[see Figs. 5 and 6 of][for an illustration]{pycs}.

\subsection{Application to RX~J1131$-$1231}

\begin{figure}[tb]
\resizebox{\hsize}{!}{\includegraphics{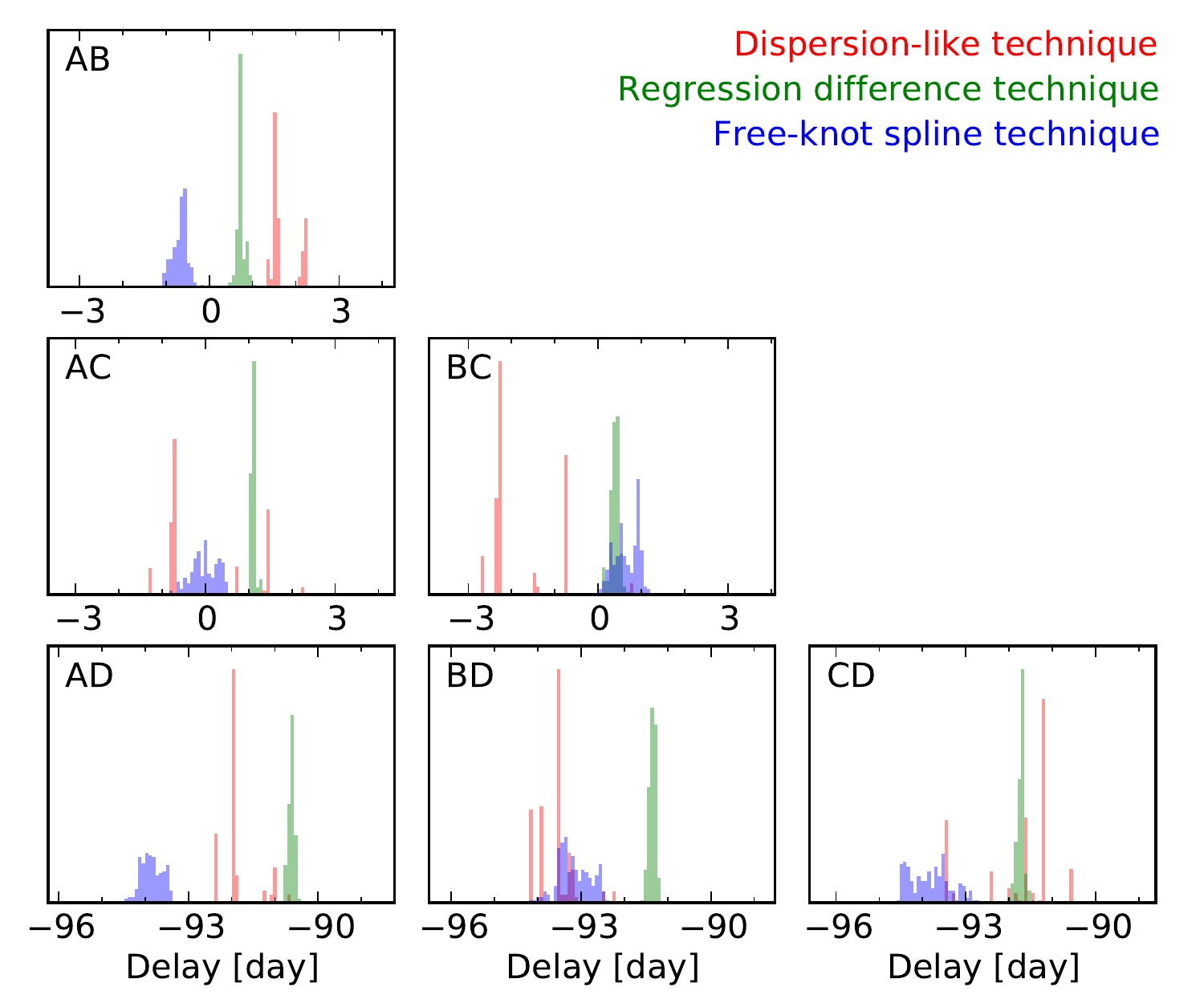}}
\caption{Analysis of the intrinsic variance of the time-delay point estimators, found by running the three curve-shifting techniques 200 times on the light curves shown in Fig. \ref{fig_lc}, starting the optimizations from random initial time shifts. The \emph{widths} of these distributions reflect the failure of the methods to converge to a single optimal solution given the data. We stress that these distributions do \emph{not} represent probability density functions of the time delays. A small intrinsic variance does not imply that a time-delay estimation is precise or accurate, only that the error surface is relatively smooth.}
\label{fig_intrinsicvariance}
\end{figure}

We begin by evaluating the robustness of the time-shift optimization of each method given the data. For this we repeatedly run the point estimators on the observed light curves. Each time, we start the optimizations from random initial conditions, using time shifts uniformly drawn $\pm 10$ days around plausible solutions. The resulting distributions of point estimates are shown in Fig. \ref{fig_intrinsicvariance}, characterizing what we call the \emph{intrinsic variance} of the estimates.

\begin{figure*}[htbp!]
\begin{center}
\vskip 20pt
\resizebox{0.8 \hsize}{!}{\includegraphics{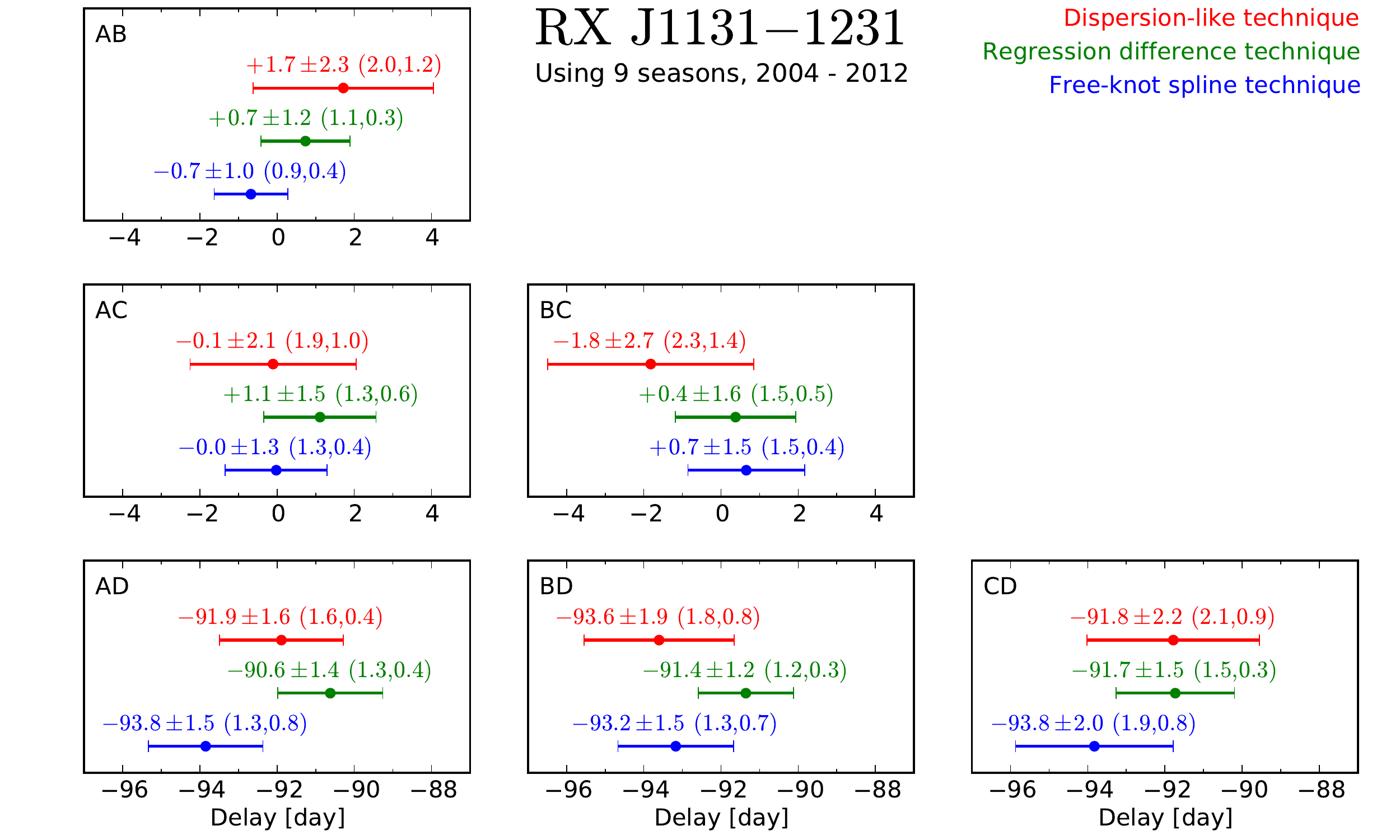}}
\caption{Time-delay measurements along with the $1\sigma$ errors for RX~J1131$-$1231 obtained by our three standard techniques from the full nine-season-long light curves shown in Fig \ref{fig_lc}. The random and systematic error contributions are given in parentheses for each delay. The error bars represent the random plus systematic errors summed in quadrature. A positive AB-delay $\Delta t_{\mathrm{AB}}$ means that image B leads image A.
We consider the measurements from the \emph{regression difference technique}, which display the lowest bias and variance in our error analysis as the delays that should be used to constrain cosmology and lens models. 
}
\label{fig_delayplot}
\vskip 30pt
\resizebox{0.8 \hsize}{!}{\includegraphics{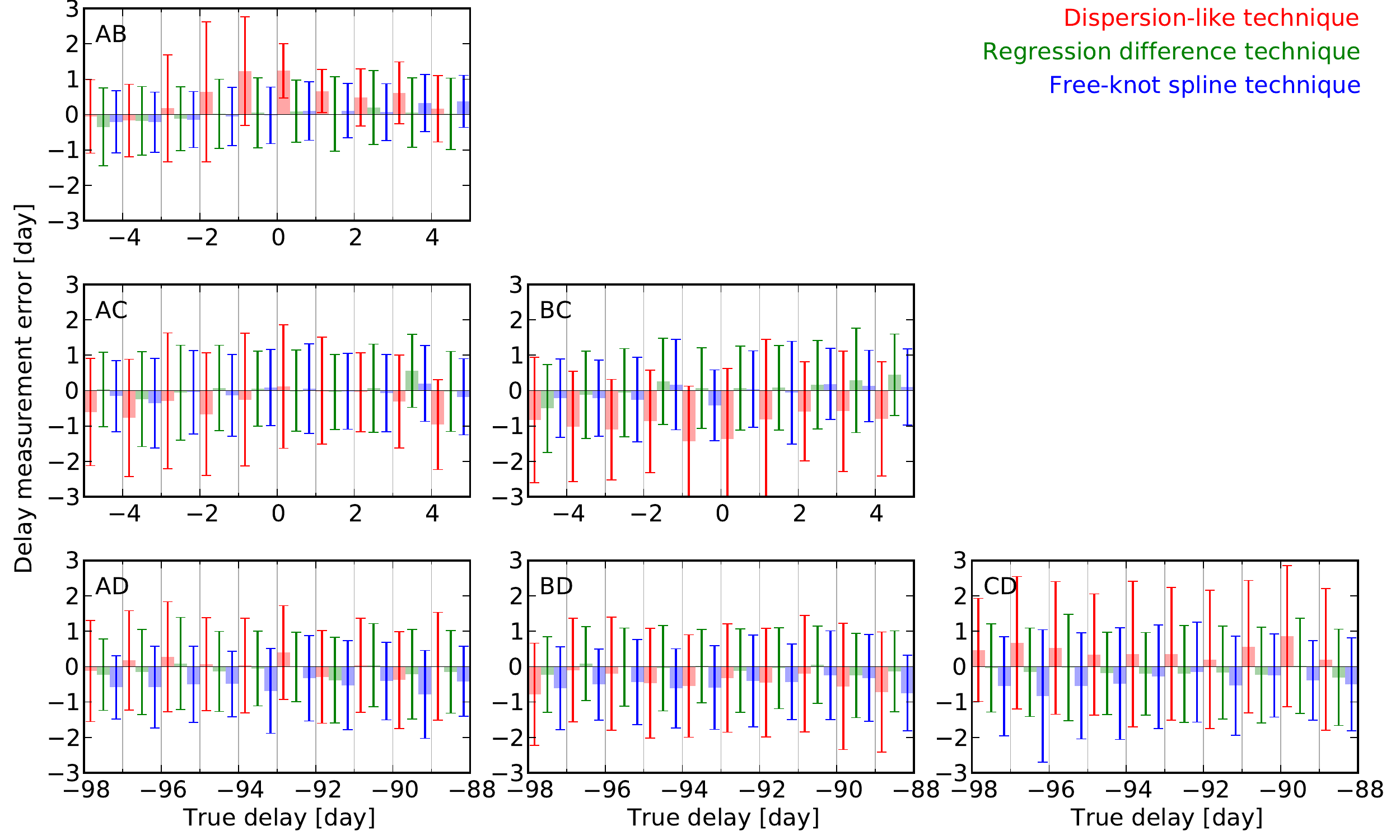}}
\caption{Results of the Monte Carlo analysis leading to the error estimates for our time-delay measurements shown in Fig. \ref{fig_delayplot}. We obtain our uncertainty estimates by applying the curve-shifting techniques to 1000 synthetic light curve sets that closely mimic the observed data but have known true time delays. The vertical axes show the delay measurement error, which is compared with the true delays used to generate the synthetic curves (horizontal axes). Separately for each panel, the outcomes are binned according to the true time delays. The bin intervals are shown as light vertical lines. Within each bin, the shaded rods and error bars show the systematic and random errors, respectively, of the delay measurements for each technique.}
\label{fig_measvstrue}
\end{center}
\end{figure*}

These tests are essential to check the reliability of the non-linear optimization algorithms for a given data set and model parameters. We stress that this procedure should not be interpreted as importance sampling of any posteriors. Generally speaking, overly flexible models dilute the time-delay information and yield higher intrinsic variances. For the free-knot spline technique we chose an average knot step of 20 days for the quasar variability spline, and 150 days for the extrinsic splines. For the dispersion-like method we model the extrinsic variability by independent linear trends on each season. If the light curves sufficiently constrain the time delays, the free-knot spline and regression difference techniques can easily be adjusted to display small intrinsic variance and roughly unimodal distributions, as in Fig. \ref{fig_intrinsicvariance}. The dispersion-like technique is inherently more sensitive to initial conditions, owing to a much higher roughness of the scalar objective function that it minimizes. As discussed in \citet{pycs}, it is important to note that smoothing the objective function does not necessarily lead to more accurate time-delay estimates. For each technique, we use the mean values of the distributions in Fig.~\ref{fig_intrinsicvariance} as our best time-delay estimates. They correspond to the points in Fig.~\ref{fig_delayplot}, which summarizes our results for RX~J1131$-$1231. 

\begin{figure}[tb]
\resizebox{\hsize}{!}{\includegraphics{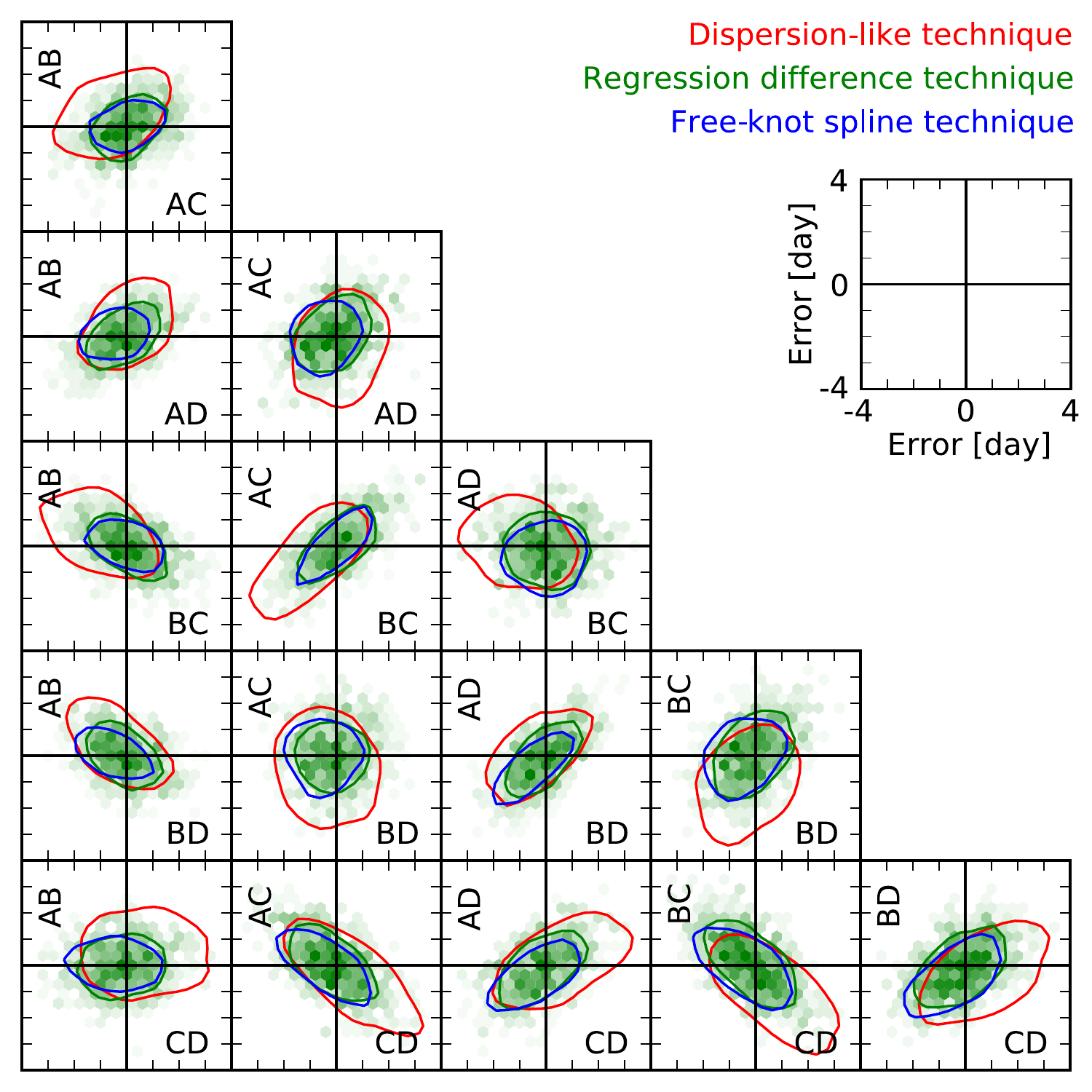}}
\caption{Correlations of delay measurement errors for synthetic light curves mimicking RX~J1131$-$1231. The measurement errors are the same as shown in Fig. \ref{fig_measvstrue}, but this time marginalizing over the true delays. Crosshairs indicate zero error, and the inset shows the scale of each panel. For clarity, only single contours at half of the maximum density are shown for two of the techniques. We observe no correlations (i.e., oblique contours) between the unrelated delay measurements along the short diagonal of this figure.}
\label{fig_covplot}
\end{figure}

The remaining part of the analysis is solely about estimating realistic error bars for these point estimates. For this we proceed by blindly running the exact same curve-shifting techniques on 1000 sets of fully synthetic light curves with true time delays randomly chosen around the measured ones. Again, we start the methods from random initial shifts. Statistics of the resulting time-delay measurement errors (i.e., measurement $-$ truth) are shown in Fig. \ref{fig_measvstrue} as a function of the true delays of the synthetic curves. In this figure, the shaded rods show the mean measurement error, which we call systematic error or bias, while the error bars indicate the standard deviation of the measurement errors, representing the random errors. The total errors for our delay measurements, shown by the error bars in Fig. \ref{fig_delayplot}, are computed by adding the maximum bias and the maximum random error of each panel in quadrature. Note that we run the curve-shifting techniques only once on each random synthetic curve set, so this error analysis takes into account the intrinsic variance. This is rather conservative, given that we use the mean result of several time-shift optimizations as our best estimates for the observed data. However, the additional contribution to the uncertainty estimates is negligible for methods with low intrinsic variance. Before discussing these results, we show in Fig. \ref{fig_covplot} the same measurement errors as in Fig. \ref{fig_measvstrue}, but plotted for each delay against each other to explore potential abnormal correlations.

\subsection{Discussion}

In terms of simple lens-model considerations, RX~J1131$-$1231 is a ``long-axis cusp lens'', whose source is located inside a cusp of the tangential caustic curve on the long axis of the potential \citep[][]{Sluse:2003cw}. Images A and D are the saddle points of the arrival-time surface, while B and C are minima \citep{Blandford:1986gn}. Figure 14 of \citet{Saha:2006kw} gives an illustration of this particular lens. Hence, the delays $\Delta t_{\mathrm AB}$ and $\Delta t_{\mathrm AC}$ are predicted to be small but positive, while $\Delta t_{\mathrm AD}$ is negative and large; the possible arrival-time orders are BCAD or CBAD.

Our measured delays, as shown in Fig. \ref{fig_delayplot}, are consistent with these predictions, although the delays between A, B, and C are compatible with being zero given the $1\sigma$ errors. \citet{Keeton:2009kc} described and quantified the exciting idea that subhalos of a lens galaxy could be detected through anomalies in observed time delays with respect to predictions from simple lens models. For a cusp lens such as RX~J1131$-$1231, these authors showed that the temporal ordering between the two minima (B and C) could easily be inverted by modest substructure. Our delay measurements do not hint at the presence of such time-delay millilensing in RX~J1131$-$1231 because they are compatible with the smooth lens models of \citet{suyu1131}.

It is reassuring to observe that the three curve-shifting methods yield consistent results for all delays despite the very different methodologies, which we see as a success of our error estimation procedure. Keep in mind, however, that the delay measurements are not independent, because they all use the same single set of observed light curves.

We now investigate in more detail the measurement-error analysis shown in Fig. \ref{fig_measvstrue}, which is based solely on synthetic curves.
Overall, we do not observe any strong dependence of a method's bias or random error on the true time delays used in the simulations. For most image pairs and methods, the random errors are more important than the almost negligible bias. Exceptions are the two results from the dispersion-like technique, which is observed to overpredict $\Delta t_{\mathrm AB}$ and underpredict $\Delta t_{\mathrm BC}$ by about one day. Remarkably, this same technique also measured the highest (lowest) delay $\Delta t_{\mathrm AB}$ ($\Delta t_{\mathrm BC}$) for the observed curves (Fig. \ref{fig_delayplot}), when compared to the other methods. We see this as another indication that the bias estimates are reliable. Finally, Fig. \ref{fig_covplot} shows no sign of abnormal correlations between measurement errors of quasar image pairs, whether they share a common quasar image or not. By construction, our time-delay uncertainty estimates for each image pair marginalize over all other pairs.

Which delay measurement method performs best on RX~J1131$-$1231 ? Given that the regression difference technique yields both the lowest biases and the smallest random errors, we simply conclude that its measurements are the most informative ones. We will use the delays from the regression difference technique, expressed with respect to quasar image B, to measure the time-delay distance toward RX~J1131$-$1231. Details and results of lens modeling, as applied to RX~J1131$-$1231, are described in detail in \citet{suyu1131}.

\subsection{Delay measurement on subsets of the light curves}
\label{subsections}

\begin{figure*}[tb]
\begin{center}
\resizebox{0.8 \hsize}{!}{\includegraphics{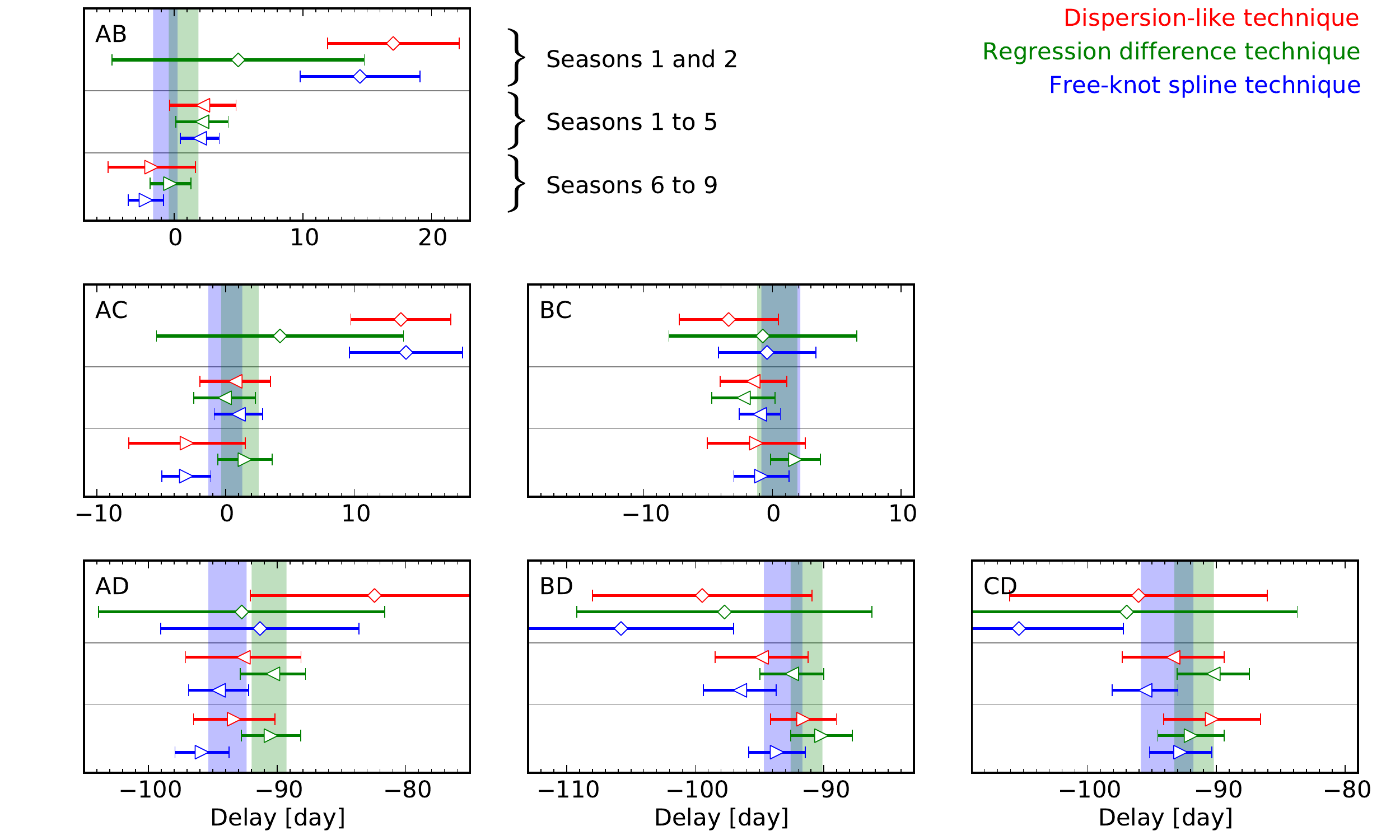}}
\caption{Application of the curve-shifting techniques to subsections of the full light curves. The square diamonds (top) show measurements using only the first two seasons, the leftward triangles (middle) use the first five seasons, and the rightward triangles (bottom) use the last four seasons. All error bars depict \emph{total} errors, as in Fig. \ref{fig_delayplot}. The blue- and green-shaded regions correspond to the interval covered by the total error bars using the full nine seasons, by the spline and regression difference techniques, respectively.}
\label{fig_seasplot}
\end{center}
\end{figure*}

Our long light curves, featuring several delay-constraining intrinsic variability patterns, suggest that we should independently measure the time delays from subsets of the observing seasons. This analysis represents an invaluable check of the consistency of the time-delay estimation procedure of \citet{pycs}, and hence for the results from the full light curves.

We analyze three subsets of the available data: (1) the first two seasons, (2) the first five seasons, and (3) the remaining last four seasons. Case (1) is chosen for comparison with \citet{Morgan:2006uw}. We perform the analyses \emph{from scratch}, without using any knowledge about the delay estimates from the full curves. In particular, we build a new model for the synthetic light curves for each case, independently adjusted so that they best resemble the observations in that time period based on the statistical criteria presented in \citet[][Section 7]{pycs}. We do not alter any parameters of the curve-shifting techniques.

Figure \ref{fig_seasplot} presents the resulting delay measurements. The data points and error bars depict the individual point estimates and the corresponding $1\sigma$ total errors obtained by each technique for the three cases. The shaded regions show the $1\sigma$ intervals from the full nine seasons taken from Fig. \ref{fig_delayplot}. We observe the following:

\begin{enumerate}

\item Using only the first two seasons, our three methods are systematically biased toward high values of $\Delta t_{\mathrm AB}$ and $\Delta t_{\mathrm AC}$. Our interpretation is that some microlensing variability in image A conspiratorially imitates a time shift, particularly around mid 2004. The estimates from the dispersion-like and spline technique roughly reproduce the results obtained from the same two seasons by \citet{Morgan:2006uw} $\Delta t_{\mathrm AB} = 12.0 \pm 1.5$, $\Delta t_{\mathrm AC} = 9.6 \pm 1.9$ and $\Delta t_{\mathrm AD} = -87 \pm 8$, but with significantly wider, yet still too low, error bars. This demonstrates the need of long monitoring programs to measure accurate time delays and minimize the influence of unfortunately placed microlensing events.

\item For these same two seasons, the regression difference method yields far better estimates for these delays -- including adequately sized error bars that encompass the delays estimated from the full light curves.
Note that the regression difference technique is the only one that does not involve an explicit \emph{model} for the microlensing.

\item For the other two divisions of the data, namely seasons 1-5 and 6-9, which are disjoint and hence independent, our methods yield consistent results. As expected, both subcases show higher error bars than the combined analysis of all seasons.

\end{enumerate}

\section{Conclusions}
\label{conclusions}

The first part of this paper describes the COSMOGRAIL data reduction procedure, which will be used to reduce all data gathered by our monitoring campaign of gravitationally lensed quasars. In the second part we apply this pipeline to our COSMOGRAIL and SMARTS observations of the quad lens RX~J1131$-$1231, leading to an unprecedented set of nine-year-long light curves of high photometric quality. Several strong and fast intrinsic quasar variability patterns constrain the time delays between the multiple images. Microlensing-related extrinsic variability is clearly present, as pointed out and analyzed in previous studies \citep[][]{Sluse:2006jv, Sluse:2007hw, Morgan:2010bw,  Dai:2010im, Chartas:2012dl}. However, this distorting signal does not prevent us from measuring accurate time delays, using the three independent algorithms of \citet{pycs}.

The best time-delay estimates of RX~J1131$-$1231 are provided by the regression difference technique. It measures the 91-day delays between D and the other quasar images to a fractional $1\sigma$ uncertainty of $1.5\%$. This error estimate is obtained by applying the techniques to synthetic curves with known time delays, which contain extrinsic variability features similar to the observed ones. We demonstrate the consistency of our error estimates by independently measuring time delays -- including error bars -- from subsets of the observed light curves of RX~J1131$-$1231. This experiment also reveals that long multi-year monitoring is essential for reliably measuring time delays, despite progress on the methods. 

The results from this paper are used to constrain the time-delay distance toward RX~J1131$-$1231 and deduce stringent implications for cosmology in \citet{suyu1131}. With our time-delay measurement errors of only 1.5\%, the accuracy of this cosmological inference is actually limited by the residual uncertainty of (1) the gravitational potential of the lens galaxy and (2) the large-scale structures along the line of sight to the quasar \citep{BarKana:1996gp}.

\begin{acknowledgements}

We are deeply grateful to the numerous observers who contributed to the data acquisition at the Swiss Euler telescope, the SMARTS 1.3-m telescope, and the Flemish-Belgian Mercator telescope. We wish to thank Sherry Suyu and Tommaso Treu for precious discussions and comments, and the anonymous referee for her/his useful suggestions. COSMOGRAIL is financially supported by the Swiss National Science Foundation (SNSF). We also acknowledge support from the International Space Science Institute in Bern, where some of this research has been discussed.
CSK and AMM are supported by NSF grant AST-1009756.
EE was partially supported by ESA and the Belgian Federal Science Policy (BELSPO) in the framework of the PRODEX Experiment Arrangement C-90312.
DS acknowledges support from the Deutsche Forschungsgemeinschaft, reference SL172/1-1.

\end{acknowledgements}

\bibliographystyle{aa} 
\bibliography{papers} 

\end{document}